\documentclass[
 aip,
 amsmath,amssymb,
 reprint,
 longbibliography
]{revtex4-2}
\usepackage[utf8]{inputenc}
\usepackage[T1]{fontenc}
\usepackage{mathptmx}
\usepackage{graphicx}
\usepackage[export]{adjustbox}
\usepackage{dcolumn}
\usepackage{bm}
\usepackage{subfigure}
\usepackage{tabularx}
\usepackage{amsmath}
\usepackage{mathrsfs}
\usepackage{hyperref}
\usepackage{booktabs}
\usepackage{color,colortbl}
\usepackage{amssymb}
\usepackage{natbib}
\usepackage{enumitem}
\usepackage{placeins}
\usepackage{CJK}
\definecolor{gr}{gray}{0.9}
\newcolumntype{S}{>{\hsize=0.3\hsize\linewidth=\hsize}X>{\hsize=1.7\hsize\linewidth=\hsize}X}
\newcommand{\rev}[1]{{\color{black}#1}}
\begin{document}
\title{Revisiting $\mathcal D^2$-law for the evaporation of dilute droplets}
\author{F. {Dalla Barba}}
\email{federico.dallabarba@unipd.it}
\affiliation{Department of Industrial Engineering, University of Padova, Padova, 35131, Italy}
\author{
\begin{CJK*}{UTF8}{gbsn}
J. Wang (王解托)
\end{CJK*}
}
\email{jietuo.wang@phd.unipd.it}
\affiliation{Centro di Ateneo di Studi e Attivit\'a Spaziali Giuseppe Colombo - CISAS, University of Padova, Padova, 35131, Italy}
\author{F. Picano}
\email{francesco.picano@unipd.it}
\affiliation{Department of Industrial Engineering, University of Padova, Padova, 35131, Italy}
\affiliation{Centro di Ateneo di Studi e Attivit\'a Spaziali Giuseppe Colombo - CISAS, University of Padova, Padova, 35131, Italy}
\begin{abstract}
In a wide range of applications, the estimate of droplet evaporation time is based on the classical $\mathcal D^2$-law, which, assuming a fast mixing and fixed environmental properties, states that the droplet surface decreases linearly with time at a determined rate. However, in many cases the predicted evaporation rate is overestimated. In this Letter, we propose a revision of the $\mathcal D^2$-law capable to accurately determine droplet evaporation rate in dilute conditions by a proper estimate of the asymptotic droplet properties. Besides a discussion of the main assumptions, we tested the proposed model against data from direct numerical simulations finding an excellent agreement for predicted droplet evaporation time in dilute turbulent jet-sprays.
\end{abstract}
\maketitle
The prediction of droplet evaporation in turbulent flows is crucial in science and applications, e.g.\ in internal combustion engines~\citep{al2015modelling} or respiratory flows~\citep{mittal2020flow}. The outbreak of COVID-19 pandemic is highly increasing the scientific relevance of this topic~\citep{chong2021extended,ng2020growth,balachandar2020}: Virus-laden droplets dispersed into the environment by an infected person spread the disease~\citep{ijerph2021}, but their dispersion range, although at the base of health guidelines, is still scientifically debated~\citep{mittal2020flow,chong2021extended,ng2020growth,balachandar2020}. For practical estimates, the prediction of droplet dispersion is still based on the work of~\citet{wells1934air}, which is founded on the $\mathcal D^2$-law to compute droplet evaporation rate. The $\mathcal D^2$-law, proposed in the seminal works of~\citet{langmuir1918evaporation} and others~\citep{spalding1950combustion,faeth1979current}, states that the surface of an evaporating droplet decreases linearly with time, at a rate fixed by the ambient properties. \rev{Although deviations from the classical $\mathcal D^2$-law have been observed in some peculiar conditions, e.g.\ for droplet sizes of the order of the mean vapor's free path~\citep{rana2019lifetime}, trans-critical evaporation of nano/micro-droplets~\citep{xiao2019molecular} and multi-component droplets~\citep{gong2021phase,nasiri2017specificity}, a linear decrease of the droplet surface is usually observed for single-component micro/millimetric droplets}. Nonetheless, recent studies on respiratory events~\citep{chong2021extended,ng2020growth} found that droplet lifetime may increase even $150$ times as compared to Wells's estimate; although a linear decrease of the droplet surface could still be observed, the evaporation rate is found to be much lower than that predicted by the $\mathcal D^2$-law. Considering that the temperature of an evaporating and isolated droplet tends to an asymptotic temperature lower than the environmental one, we propose a revision of the $\mathcal D^2$-law for \rev{single component, millimetric/micrometric droplets}. This effect, caused by the balance between heat flux and latent enthalpy, is not accounted for in the classical formulation of the $\mathcal D^2$-law, which assumes a fast mixing of the gas around the droplet. We show that, by including this effect in a revised $\mathcal D^2$-law, the droplet evaporation time can be accurately estimated in dilute conditions. To this purpose, the revised and the classical $\mathcal D^2$-laws are compared to reference data obtained from Direct Numerical Simulations (DNS) under a hybrid Eulerian-Lagrangian framework and the point-droplet approximation~\citep{weiss2020evaporating,bukhvostova2016heat,abramzon1989droplet}.
%
%
The temporal evolution of droplets smaller than the smallest flow length-scale can be Lagrangianly described using the so-called \textit{point-droplet approximation}~\citep{weiss2020evaporating,abramzon1989droplet}. Indeed, droplet temperature and radius equations read,
\begin{align}
&\frac{dT_d}{dt}=\frac{1}{3\tau_d}\left[\frac{Nu}{\bar{Pr}}\frac{\bar{c_p}}{c_l}\left(T_m-T_d\right)-\frac{Sh}{\bar{Sc}}\frac{\Delta H_v}{c_l}H_m\right],
\label{eq:1}\\
&\frac{dr^2_d}{dt}=-\frac{Sh}{\bar{Sc}}\frac{\bar{\rho}}{\rho_l}{\bar{\nu}} H_m,
\label{eq:2}
\end{align}
where $T_d$ is the droplet temperature, $r_d$ the radius and $T_m$ the temperature of the carrier mixture evaluated at the droplet center. The variable $\tau_d=2\rho_l r_d^2/(9\bar{\rho}\bar{\nu})$ is the droplet relaxation time, whereas the Schmidt and Prandtl number are $\bar{Sc}=\bar{\nu}/D_{g,v}$ and $\bar{Pr}=\bar{\nu}\bar{\rho}\bar{c_p}/\bar{k}$, with $\bar{\nu}$ the kinematic viscosity, $\bar{\rho}$ the density, $\bar{c}_p$ the isobaric specific heat capacity and $\bar{k}$ the thermal conductivity of the gas. The mass diffusivity is $D_{g,v}$, whereas $c_l$ and $\rho_l$ are the specific heat capacity and density of the liquid, and $\Delta H_v$ is the latent heat of vaporization (over-bars refer to Eulerian quantities evaluated at droplet centers). \rev{In the model a fast conductivity is implicitly assumed, i.e.\ the liquid thermal conductivity is assumed to be much higher than that of the carrier gas (low Biot number~\citep{bukhvostova2014comparison}), such that the temperature inside a droplet is uniform, but time-varying.} In analogy with $(T_m-T_d)$, which is the forcing term for convective heat transfer, the term $H_m$ drives the mass transfer rate:
\begin{align}
H_m=ln\left(\frac{1-Y_{v,m}}{1-Y_{v,d}}\right),
\label{eq:5}
\end{align}
with $Y_{v,m}(\chi_{v,m})$ the vapor mass fraction evaluated at the droplet center and $Y_{v,d}(\chi_{v,d})$ the mass fraction of a saturated gas-vapor mixture evaluated at the temperature of the droplet, $T_d$. The mass fractions depend on the vapor molar fractions which, in turn, are related to the vapor pressure as:
\begin{align}
&\chi_{v,m}=RH_m\ \frac{p_{sv}(T_m)}{p_a}, \quad \chi_{v,d}=1.0\ \frac{p_{sv}(T_d)}{p_a},
\label{eq:6}
\end{align}
where $p_a$ is the ambient pressure, $p_{sv}(T)$ the saturated vapor pressure evaluated at $T$, and $RH_m$ the relative humidity of the moist carrier evaluated at the point-droplet center. The saturated vapor pressure, $p_{sv}(T)$, depends on temperature and pressure and can be calculated by using the Clausius-Clapeyron equation. Finally, the Nusselt and Sherwood numbers are correlated to the droplet Reynolds number via the Fr\"ossling correlations, see e.g.~\citet{wang2021} for details.
\par
Based on Eq.~\eqref{eq:1}-~\eqref{eq:2},  for an isolated droplet in an environment at temperature $T_m=T_a$ and relative humidity $RH_m=RH_a$, the classical formulation of $\mathcal D^2$-law~\citep{langmuir1918evaporation} can be derived by assuming a fast mixing of the droplet atmosphere with the environment. Under this hypothesis, the quantities regulating the evaporation of the droplet are set by the bulk thermodynamic state of the ambient ($\bar{\rho}=\rho_a$, $\bar{\nu}=\nu_a$, $\bar{Sc}=Sc_a$ and $\bar Pr=Pr_a$). With the droplet temperature fixed to $T_d=T_a$, Eq.~\eqref{eq:2} can be exactly integrated as:
\begin{align}
&r_d^2=r_{d,0}^2-K\ t,
\label{eq:13}
\end{align}
where $r_{d,0}$ is the initial droplet radius and $K$ is a function of the bulk properties of the ambient $T_a$, $p_a$ and $RH_a$:
\begin{align}
K=\frac{\rho_a}{\rho_l}\frac{Sh}{Sc_a}{\nu_a}H_m(T_a,p_a,RH_a).
\label{eq:14}
\end{align}
Then, the droplet evaporation time can be estimated as $t_{d,e}={r_{d,0}^2}/{K}$. Eq.~\eqref{eq:13} is historically referred to as $\mathcal D^2$-law since it predicts a quadratic temporal evolution of the droplet diameter. It assumes that the droplet temperature is fixed and equal to the environmental one, $T_d\simeq T_a$; in turn, this condition implicitly assumes that the convective heat transfer term in the droplet temperature equation, Eq.~\eqref{eq:1}, dominates the latent heat term.
%
\begin{figure}[t!]
\centering
\includegraphics[width=1.0\linewidth]{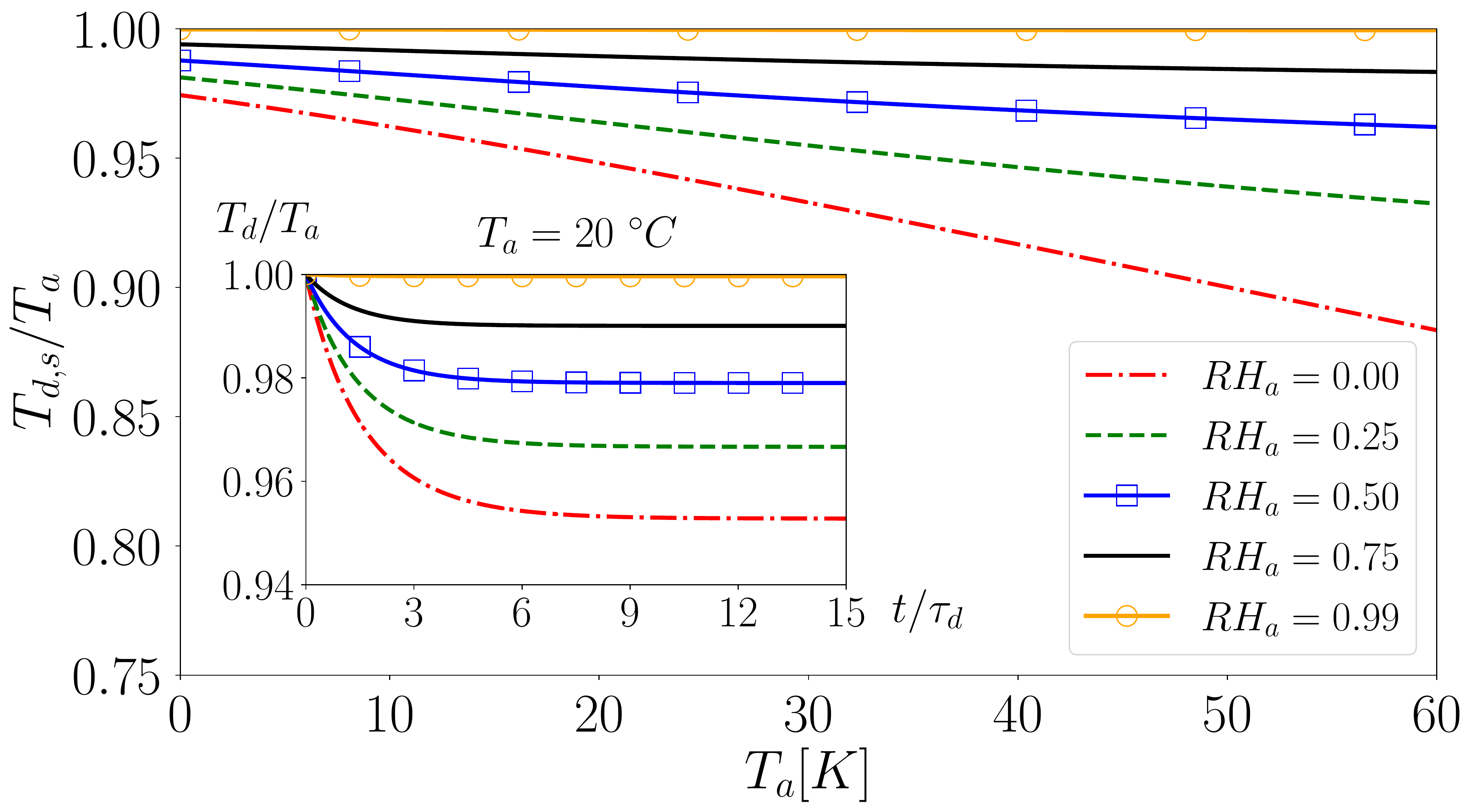}
\caption{Ratio between the droplet asymptotic temperature, $T_{d,s}$, and ambient one, $T_a$, versus ambient temperature for isolated water droplets. Inset: temporal evolution of the temperature of evaporating water droplets; the time is normalized by $\tau_{d}$. Data are obtained by the numerical solution of Eq.~\eqref{eq:1} and Eq.~\eqref{eq:2}.}
\label{fig:Td_vs_Ta}
\end{figure}
%
Nonetheless, in real cases, this condition is valid only for a short time, whose duration scales with the droplet relaxation time as sketched in the inset of Fig.~\ref{fig:Td_vs_Ta}. After a time $t\simeq6\tau_{d,0}$, the droplet temperature sets to a constant value, $T_{d,s}$, which is lower than the environmental temperature, $T_a$. The ratio between $T_a$ and $T_{d,s}$, determined by the balance between convective heat transfer and latent heat absorption, depends on the ambient temperature itself and relative humidity, as shown in Fig.~\ref{fig:Td_vs_Ta}. The droplet temperature is similar to the ambient one only for $t\ll \tau_{d,0}$, whereas, for the major part of the vaporization process, it keeps closer to the asymptotic temperature, $T_{d,s}$. Since the prediction of the droplet lifetime using the $\mathcal D^2$-law is highly sensitive to the droplet temperature, the latter may induce a significant overestimation of the evaporation rate~\citep{chong2021extended,wang2021}. Hence, we propose to revise the  $\mathcal D^2$-law by assuming a constant droplet temperature equal to the droplet asymptotic temperature and not to the ambient one: $T_d\simeq T_{d,s}$. Under these conditions, Eq.~\eqref{eq:1} and Eq.~\eqref{eq:5} become:
\begin{align}
& T_{d,s}+\frac{Pr_a}{Sc_a}\frac{Sh}{Nu}\frac{\Delta H_v}{c_{p,a}}H_m(T_a,p_a,RH_a,T_{d,s})-T_a=0,
\label{eq:16}\\
& H_m=ln\left(\frac{1-Y_{v,a}(T_a,p_a,RH_a)}{1-Y_{v,d}(T_{d,s},p_a)}\right),
\end{align}
which can be easily solved for $T_{d,s}$. The latter can be employed to compute a revised decay constant, $K_r$,
\begin{align}
&K_r=\frac{\rho_a}{\rho_l}\frac{Sh}{Sc_a}{\nu_a}H_m(T_a,p_a,RH_a,T_{d,s}),
\label{eq:rev_k}\\
&r_d^2=r_{d,0}^2-K_r\ t,
\label{eq:rev_d2l}
\end{align}
leading to a revised evaporation time, $t_{d,e}={r_{d,0}^2}/{K_r}$.
%
\begin{figure}[t!]
\centering
\includegraphics[width=1.0\linewidth]{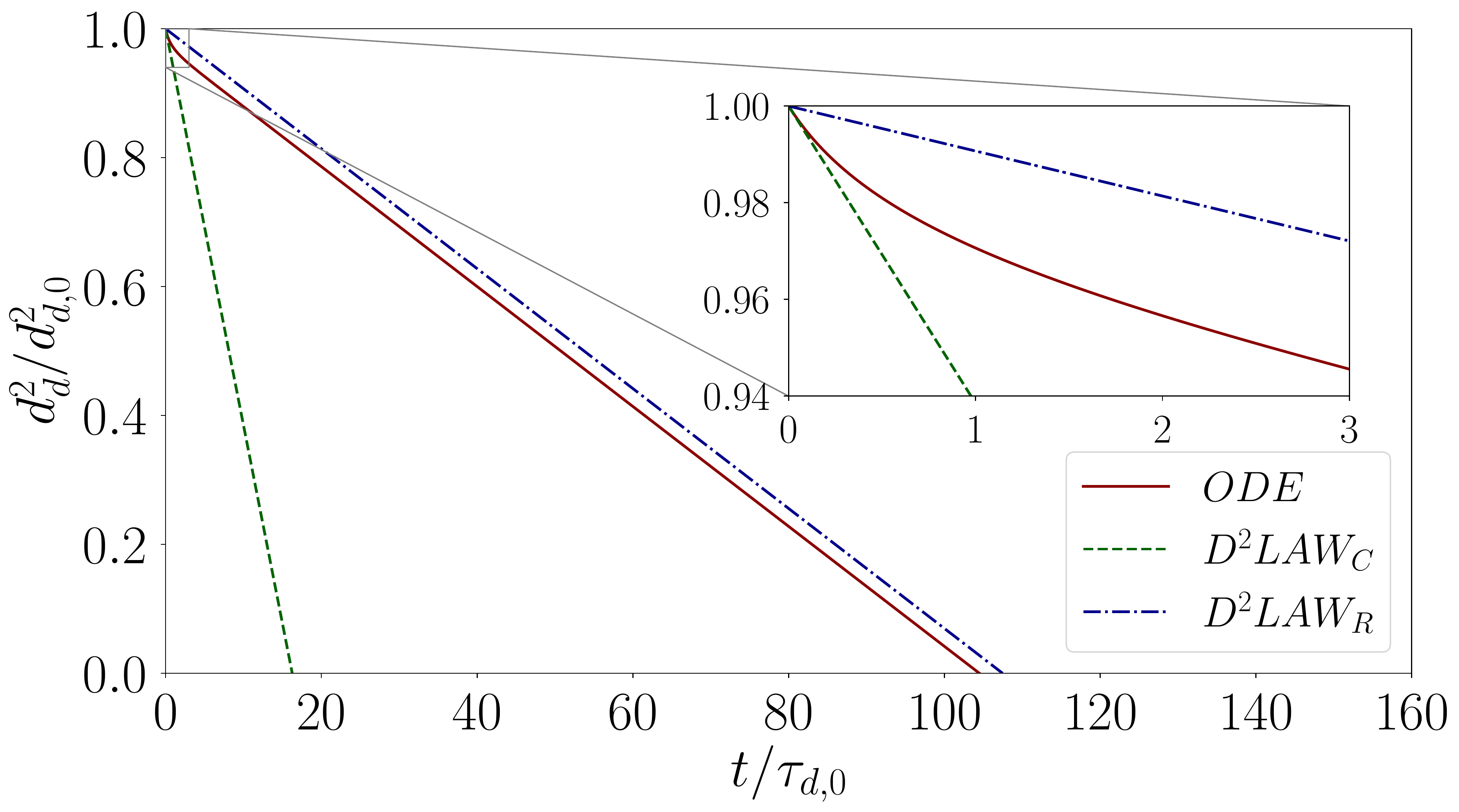}
\caption{Temporal evolution of the square diameter, $d^2_d/d^2_{d,0}$, of a water droplet in a quiescent environment at temperature $T_a=50\ ^\circ C$ and relative humidity $RH_a=0$, with $d_{d,0}$ the initial droplet diameter. Exact solution of Eq.~\eqref{eq:1} and Eq.~\eqref{eq:2} (ODE), classical $\mathcal D^2$-law ($D^2LAW_C$) and revised formulation ($D^2LAW_R$).}
\label{fig:rd_vs_t}
\end{figure}
%
Fig.~\ref{fig:rd_vs_t} shows the temporal evolution of the square diameter of a water droplet in a quiescent environment at $T_a=50\ ^\circ C$ and $RH_a=0$, providing the numerical solution of Eq.~\eqref{eq:2} and  the predictions by the classical and revised $\mathcal D^2$-laws. In the initial evaporation phase, $t\ll \tau_{d,0}$, since $T_d\simeq T_a$, the vaporization rate predicted by the classical $\mathcal D^2$-law is close to the reference one (numerical solution of Eq.~\eqref{eq:2}). Nonetheless, for $t\gg\tau_{d,0}$, the behavior predicted using the revised $\mathcal D^2$-law guarantees a much better agreement with the reference data.
%
\begin{figure}[b!]
\centering
\includegraphics[width=1.0\linewidth]{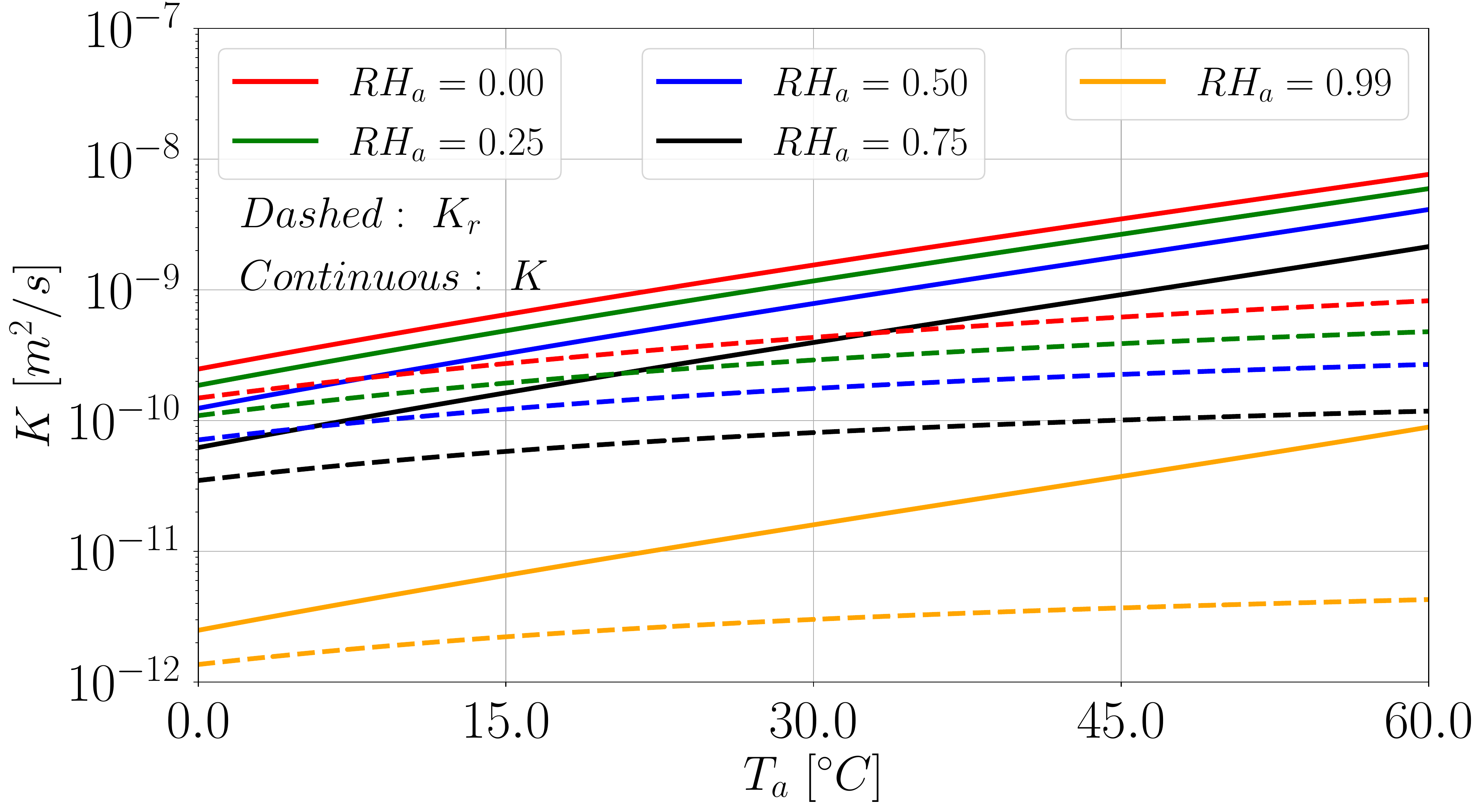}
\caption{Constant $K$ (Eq.~\eqref{eq:14}; continuous line) and $K_r$ (Eq.~\eqref{eq:rev_k}; dashed lines) for water droplets versus ambient temperature, $T_a$, for different values of the relative humidity, $RH_a$.}
\label{fig:K_vs_Ta}
\end{figure}
%
Fig.~\ref{fig:K_vs_Ta} shows the value of the decay constants $K$ and $K_r$ for water droplets versus ambient temperature at different relative humidities. The relative difference between $K_r$ and $K$ increases with $T_a$ and reduces with $RH_a$. It is worth remarking that this difference spans over nearly one order of magnitude, leading to a significant mismatch in the evaporation rates estimated by using the bulk ambient temperature or the droplet asymptotic temperature.
\par
%
%
To test the proposed revision of the $\mathcal D^2$-law, we consider a turbulent evaporating jet-spray. The well-established framework of the point-droplet equations is used to predict the temporal evolution of droplet temperature and radius (Eq.~\ref{eq:1}-Eq.~\ref{eq:2}) together with the Lagrangian equations of motion~\citep{wang2021,weiss2020evaporating,dalla2018clustering}. The droplet equations are coupled with the low-Mach number asymptotic expansion of the Navier-Stokes equations, which are directly solved to reproduce the dynamics of the carrier flow. A fully developed turbulent jet, generated from a companion DNS of a turbulent pipe flow~\citep{dalla2018clustering} at $Re_0=2U_0\,R_0/\nu=6000$, is injected into an open, quiescent environment ($U_0=9.3\, m/s$ and $R_0=4.9\cdot 10^{-3}\, m$ the pipe bulk velocity and radius). Both the jet and the environment gases consist of dry air at temperature $T_a=20^\circ C$. The computational domain is a cylinder extending for $2\pi \times 22R_0 \times 70R_0$ in the azimuthal, $\theta$, radial, $r$ and axial, $z$, directions and is discretized by a staggered grid of $N_\theta \times N_r \times N_z = 128 \times 223 \times 640$ nodes. A convective boundary condition is prescribed at the outlet section and a traction-free boundary condition is imposed on the domain side boundary to allow entrainment. A monodisperse population of water droplets of initial radius $r_{d,0}=4\mu m$ and temperature $T_{d,0}=T_a=20^\circ C$ is randomly distributed over the inflow section at each time-step. Four simulations have been performed changing the liquid mass fraction at the inflow: $\psi_0=m_l/m_g=0.0007;\,0.0028;\,0.0058;\,0.01$, with $m_l$ and $m_g$ the mass of the liquid and gas injected per unit time. \rev{The latter correspond to $8.7\times10^{-7};\,3.5\times10^{-6};\,7.0\times10^{-6};\,1.2\times10^{-5}$ in volume fraction, respectively}. Two additional simulations were conducted suppressing the numerical integration of equation Eq.~\eqref{eq:1} and Eq.~\eqref{eq:2}: In the former, the classical formulation of the $\mathcal D^2$-law, Eq.~\eqref{eq:13}, is used to evolve the droplet radius; in the latter, the revised $\mathcal D^2$-law is used, Eq.~\eqref{eq:rev_d2l}. In all cases, the position and velocity of droplets are numerically integrated according to the standard equation of point-droplet~\citep{wang2021,dalla2018clustering}. For a detailed description of the methodology and tests, the reader is referred to references~\citep{wang2021,ciottoli2020direct,dalla2018clustering}. \rev{It is worth remarking that droplet number density, mass and volume fractions are related parameters to characterize the spray dilution, governing the droplet collision/coalescence rate, mutual interaction via emitted vapor and droplet momentum back-reaction. Since the mass fraction is the parameter that mainly controls the vapor concentration and determines the balance between heat and latent enthalpy fluxes during evaporation, we selected this parameter to characterize the spray dilution in the present study.}
%
\begin{figure}[t!]
\centering
\includegraphics[width=1.0\linewidth,valign=c]{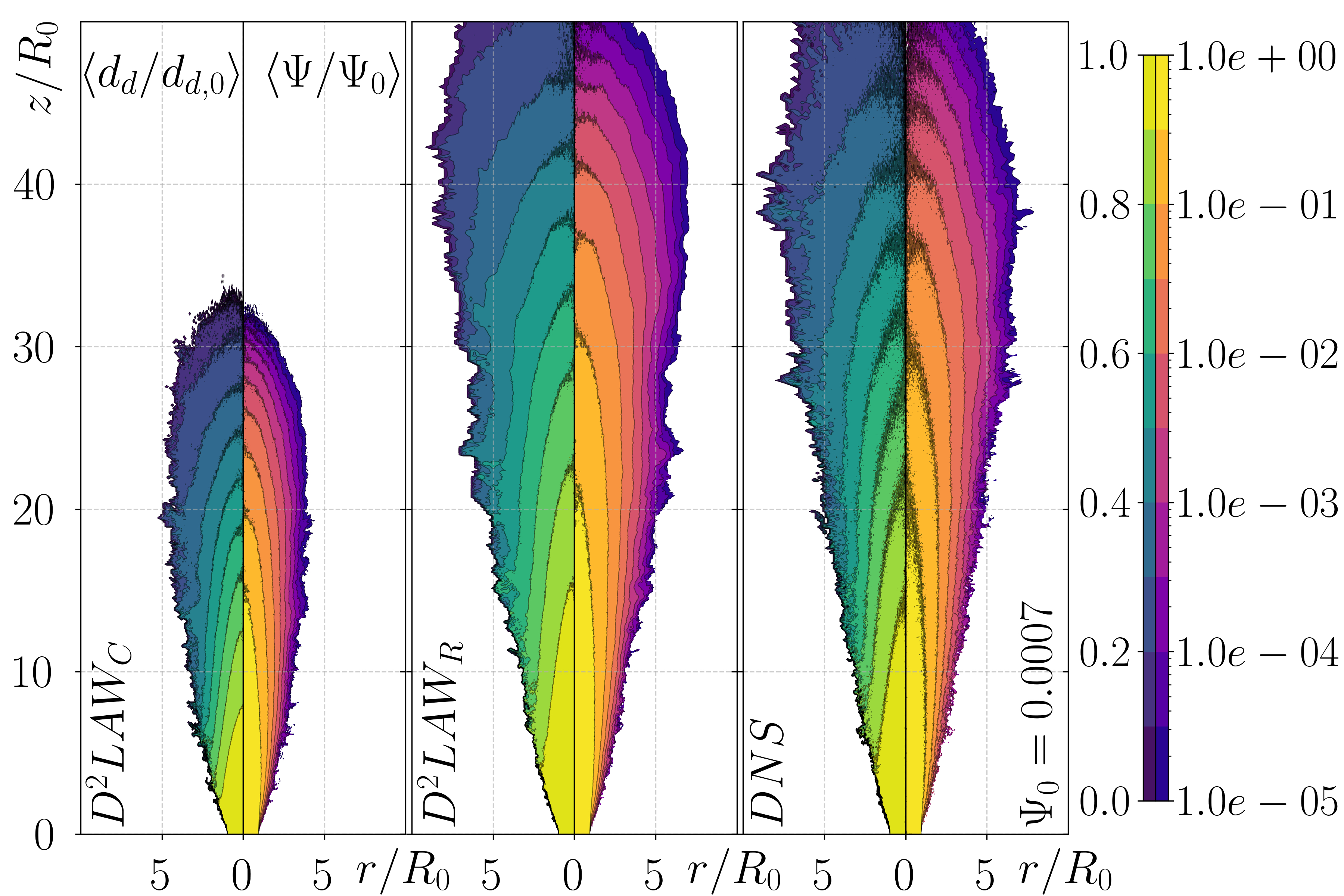}
\caption{Contour of the mean liquid mass fraction (right-half) and mean droplet diameter (left-half) for $\psi_0=0.0007$.}
\label{fig:h2o_avg}
\end{figure}
%
Fig.~\ref{fig:h2o_avg} provides the contours of the normalized mean liquid mass fraction and mean droplet diameter in the spray for the simulations based on the classical $\mathcal D^2$-law, the revised one, and on the full model for the most dilute case, $\psi_0=0.0007$. The average is computed considering a time range of $100 t_0$ ($200$ independent snapshots) after the establishment of a statistically steady regime~\citep{dalla2018clustering}, with $t_0=R_0/U_0$ the jet advection time scale. Both the distributions of the mean mass fraction and diameter computed using the revised $\mathcal D^2$-law model are in excellent agreement with the reference data. Besides, the classical $\mathcal D^2$-law results in a much shorter spray evaporation length, due to the overestimation of the vaporization rate.
The evaporation length of a turbulent jet-spray, $z_e$, can be defined as the distance from the inflow where the $99\%$ of the injected liquid mass has evaporated. This distance corresponds to a mass fraction level of $\psi/\psi_0=10^{-2}$ and, hence, to a mean droplet radius $r_{d,99}/r_{d,0}\simeq (\psi/\psi_0)^{1/3}\simeq10^{-2/3}$. By neglecting possible two-way coupling effects, an estimate of the evaporation length can be obtained by considering the self-similar behavior of the mean jet centerline velocity, $U_{z,c}$ and supposing that the droplet mean axial velocity is approximately $u_{z,d}\simeq U_{z,c}$,
\begin{align}
\frac{U_{z,c}}{U_0}=\frac{2B}{\frac{z}{R_0}-\frac{z_0}{R_0}} \quad \implies \quad \frac{R_0}{U_0}\frac{d}{dt}\left(\frac{z_d}{R_0}\right)=\frac{2B}{\frac{z_d}{R_0}-\frac{z_0}{R_0}}, 
\label{eq:res_2}
\end{align}
with $B\simeq 6$ a universal constant, $z_0$ the so-called jet virtual origin~\citep{dalla2018clustering} and $z_d$ the droplet position along the jet axis. Then, by assuming a vanishing virtual origin, $z_0/R\simeq 0$, and considering that, according to Eq.~\eqref{eq:13} and Eq.~\eqref{eq:rev_d2l}, the time required for a droplet to reduce its radius from $r_{d,0}$ to $r_{d,99}$ is $t_{d,e}=(r_{d,0}^2-r_{d,99}^2)/K$, the integration of Eq.~\eqref{eq:res_2} along $z$ leads to:
\begin{align}
\frac{z_e}{R_0}\simeq 2\sqrt{  \frac{B}{\frac{K}{R_0U_0}}\left[\left(\frac{r_{d,0}}{R_0}\right)^2-\left(\frac{r_{d,99}}{R_0}\right)^2\right]  },
\label{eq:res_3}
\end{align}
where $K$ is either the decay rate defined according to Eq.~\eqref{eq:14} or $K_r$ defined in Eq.~\eqref{eq:rev_k}. By using Eq.~\eqref{eq:res_3}, the resulting evaporation lengths are $z_e/R_0\simeq 28$ and $z_e/R_0\simeq 46$, computed by using $K$ and $K_r$, respectively. These values are consistent with the axial locations where the $99\%$ of the injected liquid mass has evaporated according to Fig.~\ref{fig:h2o_avg} ($\psi/\psi_0\simeq10^{-2}$) showing the superior estimate of the proposed model.
%
\begin{figure}[t!]
\centering
\includegraphics[width=1.0\linewidth,valign=c]{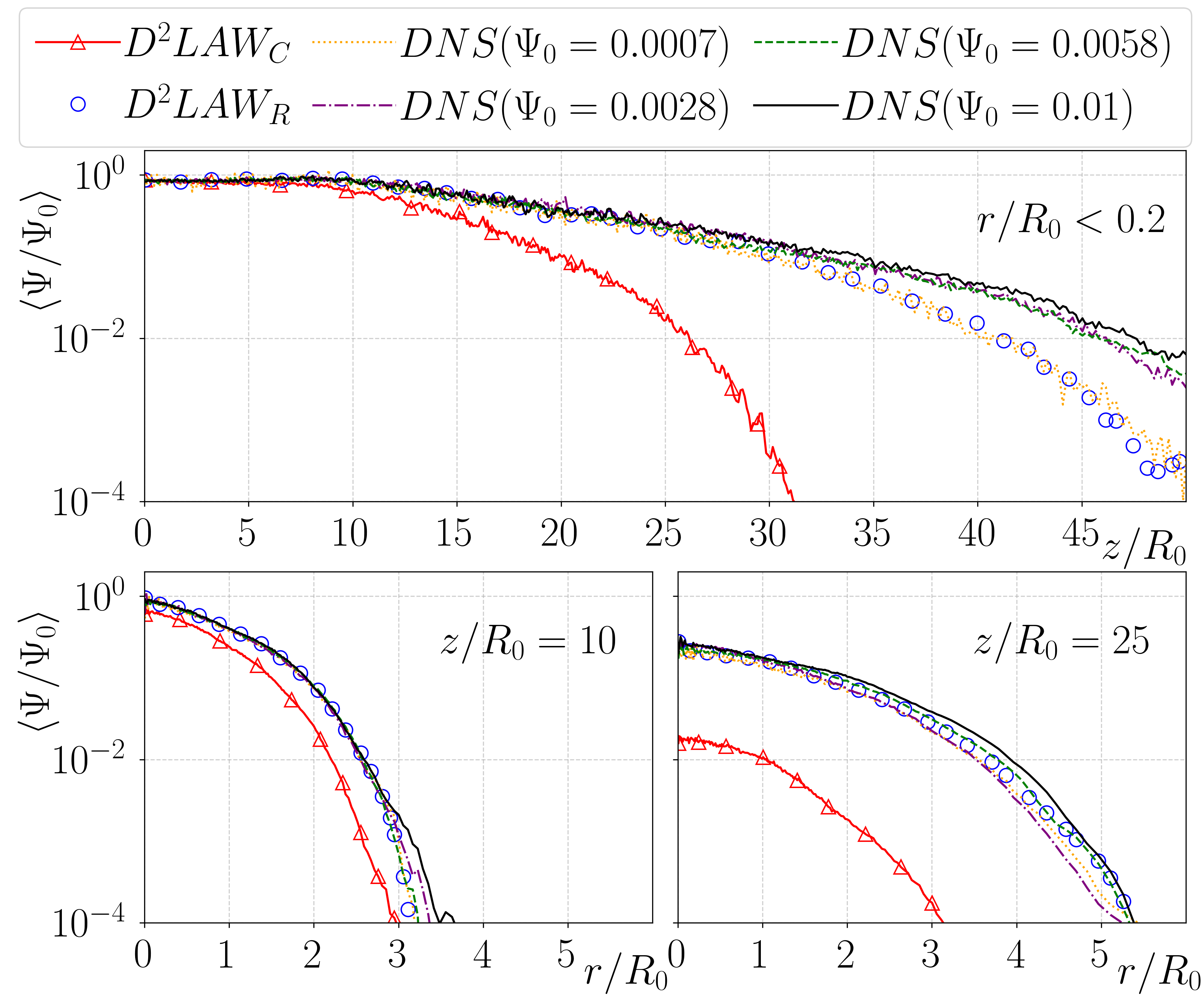}
\caption{Upper panel: mean liquid mass fraction along the centerline of the jet computed for $r/R_0<0.2$. Lower panels: mean liquid mass fraction along the radial direction for two different distances from the inflow, $z/R_0=10$ (left) and $z/R_0=25$ (right).} 
\label{fig:h2o_lines}
\end{figure}
%
To better quantify the evaporation length, the upper panel of Fig.~\ref{fig:h2o_lines} provides the mean liquid mass fraction computed along the jet centerline. In the near-field, up to $10$ jet radii away from the inlet, all the curves are close to each other. In this region the actual droplet temperature is close to the environmental one, $T_d\simeq T_a$; being the droplet and ambient temperature similar, the actual evaporation rate is close to the one imposed by the environmental conditions. Moving downstream, the droplet temperature decreases and the classical $\mathcal D^2$-law underestimates the mean liquid mass fraction, since droplets evaporate at a constant rate driven by the ambient temperature. On the other hand, the simulation based on the revised $\mathcal D^2$-law better approximates the actual droplet temperature, providing a more accurate estimate of the mean evaporation rate and axial evolution of the mean mass fraction. For the latter, the agreement with all the reference simulations is excellent, except for the far-field of the higher mass fraction cases, $\psi_0>0.0028$. In non-dilute cases, the mutual droplet interactions, via the emitted vapor, cause a deviation of the local evaporation rate from the one predicted by the $\mathcal D^2$-laws. Similar results are obtained for the radial profiles of the mean mass fraction provided in the lower panel of Fig.~\ref{fig:h2o_lines} and computed at $z/R_0=10$ and $z/R_0=25$. At $z/R_0=10$ all curves are still similar, although the simulation employing the classical $\mathcal D^2$-law shows differences with respect to reference data. Moving downstream, the underestimation of the liquid mass fraction resulting from the classical $\mathcal D^2$-law is more apparent, whereas the revised $\mathcal D^2$-law still provides an excellent estimate.
\par
%
\begin{figure}[t!]
\centering
\includegraphics[width=1.0\linewidth,valign=c]{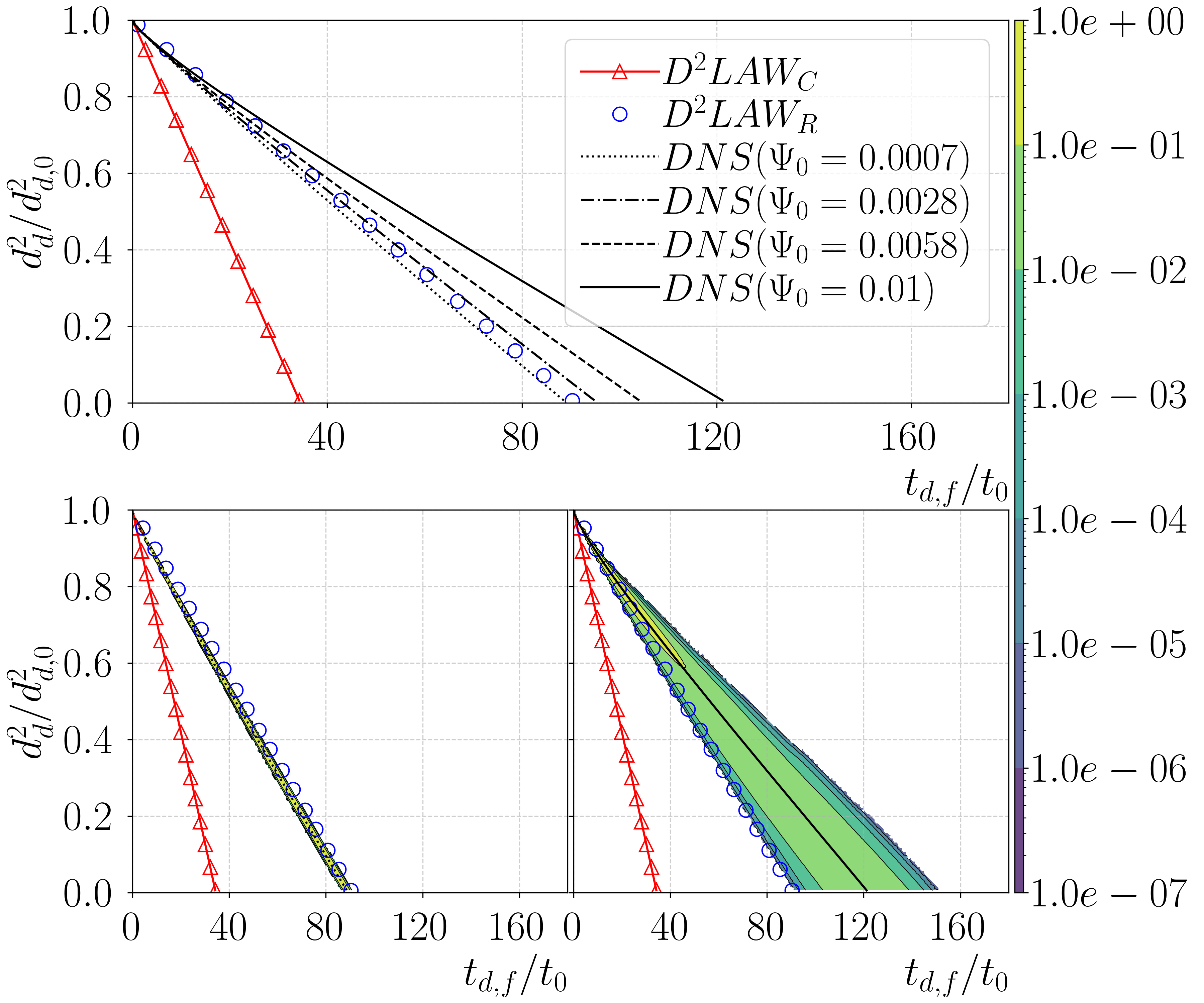}
\caption{Upper panel: mean droplet square diameter, $d_{d}^2/d^2_{d,0}$ versus mean droplet flight time, $t_{d,f}/t_0$. Lower panel: joint probability density functions of the droplet square diameter versus flight time for $\psi_0=0.0007$ (left) and $\psi_0=0.01$ (right).}
\label{fig:h2o_jpdf}
\end{figure}
%
Considering Lagrangian statistics, the top panel of Fig.~\ref{fig:h2o_jpdf} shows the mean droplet square diameter, $d_{d}^2/d^2_{d,0}$ versus the flight time, $t_{d,f}/t_0$, calculated since the injection. In all cases, the evolution of $d_{d}^2/d^2_{d,0}$ computed from full DNSs are compared with the estimates from the classical and revised $\mathcal D^2$-laws. The predictions obtained from the proposed revised model appear much more accurate with respect to the classical formulation. For mass fraction equal, or below, to $\psi_0=0.0028$ the agreement is excellent, whereas the classical $\mathcal D^2$-law predicts an evaporation time shorter than the half with respect to the reference DNS. Increasing the injected mass fraction, the mutual interactions of droplets slow down the evaporation. From the figure, we conclude that the proposed revised $\mathcal D^2$-law optimally describe droplet evaporation in dilute conditions up to mass fractions of the order of $\psi_0\sim10^{-3}$. The effect of the mutual interactions of the evaporating droplets is better highlighted in the lower panels of Fig.~\ref{fig:h2o_jpdf}, which provides the Joint Probability Density Functions (JPDF) of $d_{d}^2/d^2_{d,0}$ versus $t_{d,f}/t_0$ for the lowest and highest mass fraction cases, $\psi_0=0.0007$ and $\psi_0=0.01$. It is worth noting that, in the most dilute conditions the JPDF is superimposed on the straight line determined by the revised $\mathcal D^2$-law: Each droplet evaporates as an isolated one, which is perfectly described by the proposed model. On the contrary, for the highest mass fraction, the JPDF spreads over a wider range and the evaporation times are longer than that predicted by the proposed model (around $35\%$ in mean). This indicates that droplets exhibit different Lagrangian histories and, as expected, the proposed model cannot account for this complex interaction, typical of non-dilute conditions.
\par
%
\begin{figure}[t!]
\centering
\includegraphics[width=1.0\linewidth,valign=c]{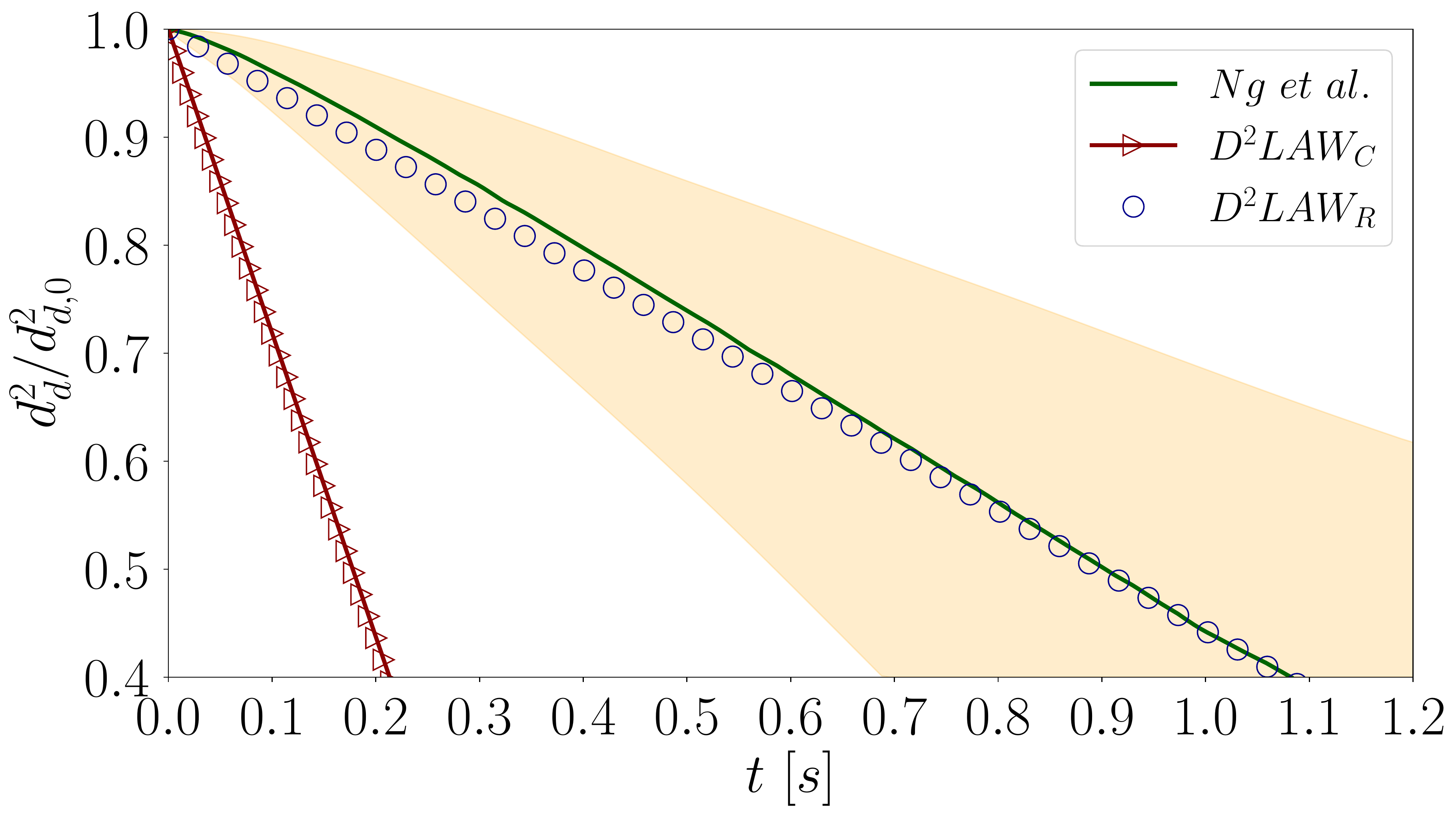}
\caption{Mean droplet square diameter, $d_{d}^2/d^2_{d,0}$ versus flight time. Green line data, by~\citet{ng2020growth}; the highlighted area represents the  distribution range  of the droplet diameters observed.}
\label{fig:h2o_lohse}
\end{figure}
%
To highlight the generality of the present study, we employed our model against the dataset from recent independent works by~\citet{ng2020growth} and~\citet{chong2021extended} on respiratory droplets evaporation. We focus on their numerical simulation on droplets expelled in a turbulent respiratory puff (cough) at $T_a=30^\circ C$ and $RH_a=90\%$. The authors observed a linear decrease of the droplet surface but found an evaporation rate much lower than that estimated by the classical $\mathcal D^2$-law. In Fig.~\ref{fig:h2o_lohse} we compare the mean droplet square diameter, $d_{d}^2/d^2_{d,0}$, versus mean droplet flight time obtained by~\citet{ng2020growth} with that predicted by using the classical and the revised $\mathcal D^2$-laws. Our revised formulation accurately reproduces the temporal evolution of mean square droplet diameter, whereas the classical prediction strongly overestimates the vaporization rate, leading to shorter evaporation times. It is worth noting that the additional cases presented in~\citet{ng2020growth} pertain to conditions where the emitted gaseous jet strongly differs from the environmental ones, so the present model cannot be directly applied.\par \rev{Summarizing the applicability limits of the present model, the droplet size should be smaller than the smallest flow length scales (point-droplet approximation) and well above the non-continuum length-scales, i.e.\  nano-droplets~\citep{rana2019lifetime}. Hence, we can state the model is applicable for droplet size  below millimeters and above a micrometer.} \rev{In addition, beyond a dilute mass fraction ($\psi_0<10^{-2}$), the basic assumptions behind the presented revision of the $\mathcal D^2$-law requires  the thermodynamic ambient conditions to be sufficiently homogeneous and similar to that of the gaseous jet. These assumptions are also at the base of the classical $\mathcal D^2$-law.}\par
%
%
\rev{To conclude}, the present study demonstrates that reformulating the $\mathcal D^2$-law by using the asymptotic temperature of isolated, evaporating droplets instead of the ambient one, allows a superior description of \rev{the evaporation of dilute, millimetric/micrometric and single component droplets} in turbulent flows with respect to the classical $\mathcal D^2$-law. An excellent agreement of the predictions by our revised model against reference DNS is found for jet-sprays up to a liquid mass fraction of the order of $\psi_0\sim10^{-3}$, whereas the agreement is still acceptable up to $\psi_0\sim10^{-2}$. The authors believe that the proposed revision of the $\mathcal D^2$-law will contribute to improve practical estimates of droplet evaporation times. In the context of respiratory droplet dispersion, we expect that present findings could be significant for a revision of the Wells theory~\citep{wells1934air}, which is based on the classical $\mathcal D^2$-law.\\

\noindent\emph{Acknowledgement.} J.W. is grateful to the China Scholarship Council (CSC) for supporting this research (201806250023).\\

\bibliography{biblio}

\begin{thebibliography}{21}%
\makeatletter
\providecommand \@ifxundefined [1]{%
 \@ifx{#1\undefined}
}%
\providecommand \@ifnum [1]{%
 \ifnum #1\expandafter \@firstoftwo
 \else \expandafter \@secondoftwo
 \fi
}%
\providecommand \@ifx [1]{%
 \ifx #1\expandafter \@firstoftwo
 \else \expandafter \@secondoftwo
 \fi
}%
\providecommand \natexlab [1]{#1}%
\providecommand \enquote  [1]{``#1''}%
\providecommand \bibnamefont  [1]{#1}%
\providecommand \bibfnamefont [1]{#1}%
\providecommand \citenamefont [1]{#1}%
\providecommand \href@noop [0]{\@secondoftwo}%
\providecommand \href [0]{\begingroup \@sanitize@url \@href}%
\providecommand \@href[1]{\@@startlink{#1}\@@href}%
\providecommand \@@href[1]{\endgroup#1\@@endlink}%
\providecommand \@sanitize@url [0]{\catcode `\\12\catcode `\$12\catcode
  `\&12\catcode `\#12\catcode `\^12\catcode `\_12\catcode `\%12\relax}%
\providecommand \@@startlink[1]{}%
\providecommand \@@endlink[0]{}%
\providecommand \url  [0]{\begingroup\@sanitize@url \@url }%
\providecommand \@url [1]{\endgroup\@href {#1}{\urlprefix }}%
\providecommand \urlprefix  [0]{URL }%
\providecommand \Eprint [0]{\href }%
\providecommand \doibase [0]{https://doi.org/}%
\providecommand \selectlanguage [0]{\@gobble}%
\providecommand \bibinfo  [0]{\@secondoftwo}%
\providecommand \bibfield  [0]{\@secondoftwo}%
\providecommand \translation [1]{[#1]}%
\providecommand \BibitemOpen [0]{}%
\providecommand \bibitemStop [0]{}%
\providecommand \bibitemNoStop [0]{.\EOS\space}%
\providecommand \EOS [0]{\spacefactor3000\relax}%
\providecommand \BibitemShut  [1]{\csname bibitem#1\endcsname}%
\let\auto@bib@innerbib\@empty
\bibitem [{\citenamefont {{Al Qubeissi}}\ \emph {et~al.}(2015)\citenamefont
  {{Al Qubeissi}}, \citenamefont {Sazhin}, \citenamefont {Turner},
  \citenamefont {Begg}, \citenamefont {Crua},\ and\ \citenamefont
  {Heikal}}]{al2015modelling}%
  \BibitemOpen
  \bibfield  {author} {\bibinfo {author} {\bibfnamefont {M.}~\bibnamefont {{Al
  Qubeissi}}}, \bibinfo {author} {\bibfnamefont {S.~S.}\ \bibnamefont
  {Sazhin}}, \bibinfo {author} {\bibfnamefont {J.}~\bibnamefont {Turner}},
  \bibinfo {author} {\bibfnamefont {S.}~\bibnamefont {Begg}}, \bibinfo {author}
  {\bibfnamefont {C.}~\bibnamefont {Crua}},\ and\ \bibinfo {author}
  {\bibfnamefont {M.~R.}\ \bibnamefont {Heikal}},\ }\href
  {https://doi.org/10.1016/j.fuel.2015.06.028} {\bibfield  {journal} {\bibinfo
  {journal} {Fuel}\ }\textbf {\bibinfo {volume} {159}},\ \bibinfo {pages} {373}
  (\bibinfo {year} {2015})}\BibitemShut {NoStop}%
\bibitem [{\citenamefont {Mittal}\ \emph {et~al.}(2020)\citenamefont {Mittal},
  \citenamefont {Ni},\ and\ \citenamefont {Seo}}]{mittal2020flow}%
  \BibitemOpen
  \bibfield  {author} {\bibinfo {author} {\bibfnamefont {R.}~\bibnamefont
  {Mittal}}, \bibinfo {author} {\bibfnamefont {R.}~\bibnamefont {Ni}},\ and\
  \bibinfo {author} {\bibfnamefont {J.~H.}\ \bibnamefont {Seo}},\ }\href
  {https://doi.org/10.1017/jfm.2020.330} {\bibfield  {journal} {\bibinfo
  {journal} {J. Fluid Mech.}\ }\textbf {\bibinfo {volume} {894}},\ \bibinfo
  {pages} {F2} (\bibinfo {year} {2020})}\BibitemShut {NoStop}%
\bibitem [{\citenamefont {Chong}\ \emph {et~al.}(2021)\citenamefont {Chong},
  \citenamefont {Ng}, \citenamefont {Hori}, \citenamefont {Yang}, \citenamefont
  {Verzicco},\ and\ \citenamefont {Lohse}}]{chong2021extended}%
  \BibitemOpen
  \bibfield  {author} {\bibinfo {author} {\bibfnamefont {K.~L.}\ \bibnamefont
  {Chong}}, \bibinfo {author} {\bibfnamefont {C.~S.}\ \bibnamefont {Ng}},
  \bibinfo {author} {\bibfnamefont {N.}~\bibnamefont {Hori}}, \bibinfo {author}
  {\bibfnamefont {R.}~\bibnamefont {Yang}}, \bibinfo {author} {\bibfnamefont
  {R.}~\bibnamefont {Verzicco}},\ and\ \bibinfo {author} {\bibfnamefont
  {D.}~\bibnamefont {Lohse}},\ }\href
  {https://doi.org/10.1103/PhysRevLett.126.034502} {\bibfield  {journal}
  {\bibinfo  {journal} {Phys. Rev. Lett.}\ }\textbf {\bibinfo {volume} {126}},\
  \bibinfo {pages} {034502} (\bibinfo {year} {2021})}\BibitemShut {NoStop}%
\bibitem [{\citenamefont {Ng}\ \emph {et~al.}(2020)\citenamefont {Ng},
  \citenamefont {Chong}, \citenamefont {Yang}, \citenamefont {Li},
  \citenamefont {Verzicco},\ and\ \citenamefont {Lohse}}]{ng2020growth}%
  \BibitemOpen
  \bibfield  {author} {\bibinfo {author} {\bibfnamefont {C.~S.}\ \bibnamefont
  {Ng}}, \bibinfo {author} {\bibfnamefont {K.~L.}\ \bibnamefont {Chong}},
  \bibinfo {author} {\bibfnamefont {R.}~\bibnamefont {Yang}}, \bibinfo {author}
  {\bibfnamefont {M.}~\bibnamefont {Li}}, \bibinfo {author} {\bibfnamefont
  {R.}~\bibnamefont {Verzicco}},\ and\ \bibinfo {author} {\bibfnamefont
  {D.}~\bibnamefont {Lohse}},\ }\href@noop {} {\bibfield  {journal} {\bibinfo
  {journal} {medRxiv}\ } (\bibinfo {year} {2020})}\BibitemShut {NoStop}%
\bibitem [{\citenamefont {Balachandar}\ \emph {et~al.}(2020)\citenamefont
  {Balachandar}, \citenamefont {Zaleski}, \citenamefont {Soldati},
  \citenamefont {Ahmadi},\ and\ \citenamefont {Bourouiba}}]{balachandar2020}%
  \BibitemOpen
  \bibfield  {author} {\bibinfo {author} {\bibfnamefont {S.}~\bibnamefont
  {Balachandar}}, \bibinfo {author} {\bibfnamefont {S.}~\bibnamefont
  {Zaleski}}, \bibinfo {author} {\bibfnamefont {A.}~\bibnamefont {Soldati}},
  \bibinfo {author} {\bibfnamefont {G.}~\bibnamefont {Ahmadi}},\ and\ \bibinfo
  {author} {\bibfnamefont {L.}~\bibnamefont {Bourouiba}},\ }\href
  {https://doi.org/10.1016/j.ijmultiphaseflow.2020.103439} {\bibfield
  {journal} {\bibinfo  {journal} {Int. J. Multiph. Flow}\ }\textbf {\bibinfo
  {volume} {132}},\ \bibinfo {pages} {103439} (\bibinfo {year}
  {2020})}\BibitemShut {NoStop}%
\bibitem [{\citenamefont {Setti}\ \emph {et~al.}(2020)\citenamefont {Setti},
  \citenamefont {Passarini}, \citenamefont {{De Gennaro}}, \citenamefont
  {Barbieri}, \citenamefont {Perrone}, \citenamefont {Borelli}, \citenamefont
  {Palmisani}, \citenamefont {{Di Gilio}}, \citenamefont {Piscitelli},\ and\
  \citenamefont {Miani}}]{ijerph2021}%
  \BibitemOpen
  \bibfield  {author} {\bibinfo {author} {\bibfnamefont {L.}~\bibnamefont
  {Setti}}, \bibinfo {author} {\bibfnamefont {F.}~\bibnamefont {Passarini}},
  \bibinfo {author} {\bibfnamefont {G.}~\bibnamefont {{De Gennaro}}}, \bibinfo
  {author} {\bibfnamefont {P.}~\bibnamefont {Barbieri}}, \bibinfo {author}
  {\bibfnamefont {M.~G.}\ \bibnamefont {Perrone}}, \bibinfo {author}
  {\bibfnamefont {M.}~\bibnamefont {Borelli}}, \bibinfo {author} {\bibfnamefont
  {J.}~\bibnamefont {Palmisani}}, \bibinfo {author} {\bibfnamefont
  {A.}~\bibnamefont {{Di Gilio}}}, \bibinfo {author} {\bibfnamefont
  {P.}~\bibnamefont {Piscitelli}},\ and\ \bibinfo {author} {\bibfnamefont
  {A.}~\bibnamefont {Miani}},\ }\href {https://doi.org/10.3390/ijerph17082932}
  {\bibfield  {journal} {\bibinfo  {journal} {Int. J. Environ. Res. Public
  Health}\ }\textbf {\bibinfo {volume} {17}},\ \bibinfo {pages} {2932}
  (\bibinfo {year} {2020})}\BibitemShut {NoStop}%
\bibitem [{\citenamefont {Wells}(1934)}]{wells1934air}%
  \BibitemOpen
  \bibfield  {author} {\bibinfo {author} {\bibfnamefont {W.~F.}\ \bibnamefont
  {Wells}},\ }\href@noop {} {\bibfield  {journal} {\bibinfo  {journal} {Am. J.
  Hyg.}\ }\textbf {\bibinfo {volume} {20}},\ \bibinfo {pages} {611} (\bibinfo
  {year} {1934})}\BibitemShut {NoStop}%
\bibitem [{\citenamefont {Langmuir}(1918)}]{langmuir1918evaporation}%
  \BibitemOpen
  \bibfield  {author} {\bibinfo {author} {\bibfnamefont {I.}~\bibnamefont
  {Langmuir}},\ }\href {https://doi.org/10.1103/PhysRev.12.368} {\bibfield
  {journal} {\bibinfo  {journal} {Phys. Rev.}\ }\textbf {\bibinfo {volume}
  {12}},\ \bibinfo {pages} {368} (\bibinfo {year} {1918})}\BibitemShut
  {NoStop}%
\bibitem [{\citenamefont {Spalding}(1950)}]{spalding1950combustion}%
  \BibitemOpen
  \bibfield  {author} {\bibinfo {author} {\bibfnamefont {D.~B.}\ \bibnamefont
  {Spalding}},\ }\href {https://doi.org/10.1038/165160a0} {\bibfield  {journal}
  {\bibinfo  {journal} {Nature}\ }\textbf {\bibinfo {volume} {165}},\ \bibinfo
  {pages} {160} (\bibinfo {year} {1950})}\BibitemShut {NoStop}%
\bibitem [{\citenamefont {Faeth}(1979)}]{faeth1979current}%
  \BibitemOpen
  \bibfield  {author} {\bibinfo {author} {\bibfnamefont {G.~M.}\ \bibnamefont
  {Faeth}},\ }\href {https://doi.org/10.1016/B978-0-08-024780-9.50013-7}
  {\bibfield  {journal} {\bibinfo  {journal} {Energy Combust. Sci.}\ ,\
  \bibinfo {pages} {149}} (\bibinfo {year} {1979})}\BibitemShut {NoStop}%
\bibitem [{\citenamefont {Rana}\ \emph {et~al.}(2019)\citenamefont {Rana},
  \citenamefont {Lockerby},\ and\ \citenamefont
  {Sprittles}}]{rana2019lifetime}%
  \BibitemOpen
  \bibfield  {author} {\bibinfo {author} {\bibfnamefont {A.~S.}\ \bibnamefont
  {Rana}}, \bibinfo {author} {\bibfnamefont {D.~A.}\ \bibnamefont {Lockerby}},\
  and\ \bibinfo {author} {\bibfnamefont {J.~E.}\ \bibnamefont {Sprittles}},\
  }\href {https://doi.org/10.1103/PhysRevLett.123.154501} {\bibfield  {journal}
  {\bibinfo  {journal} {Phys. Rev. Lett.}\ }\textbf {\bibinfo {volume} {123}},\
  \bibinfo {pages} {154501} (\bibinfo {year} {2019})}\BibitemShut {NoStop}%
\bibitem [{\citenamefont {Xiao}\ \emph {et~al.}(2019)\citenamefont {Xiao},
  \citenamefont {Luo}, \citenamefont {Ma},\ and\ \citenamefont
  {Shuai}}]{xiao2019molecular}%
  \BibitemOpen
  \bibfield  {author} {\bibinfo {author} {\bibfnamefont {G.}~\bibnamefont
  {Xiao}}, \bibinfo {author} {\bibfnamefont {K.~H.}\ \bibnamefont {Luo}},
  \bibinfo {author} {\bibfnamefont {X.}~\bibnamefont {Ma}},\ and\ \bibinfo
  {author} {\bibfnamefont {S.}~\bibnamefont {Shuai}},\ }\href
  {https://doi.org/10.1016/j.proci.2018.09.020} {\bibfield  {journal} {\bibinfo
   {journal} {Proc. Combust. Inst.}\ }\textbf {\bibinfo {volume} {37}},\
  \bibinfo {pages} {3219} (\bibinfo {year} {2019})}\BibitemShut {NoStop}%
\bibitem [{\citenamefont {Gong}\ \emph {et~al.}(2021)\citenamefont {Gong},
  \citenamefont {Xiao}, \citenamefont {Ma}, \citenamefont {Luo}, \citenamefont
  {Shuai},\ and\ \citenamefont {Xu}}]{gong2021phase}%
  \BibitemOpen
  \bibfield  {author} {\bibinfo {author} {\bibfnamefont {Y.}~\bibnamefont
  {Gong}}, \bibinfo {author} {\bibfnamefont {G.}~\bibnamefont {Xiao}}, \bibinfo
  {author} {\bibfnamefont {X.}~\bibnamefont {Ma}}, \bibinfo {author}
  {\bibfnamefont {K.~H.}\ \bibnamefont {Luo}}, \bibinfo {author} {\bibfnamefont
  {S.}~\bibnamefont {Shuai}},\ and\ \bibinfo {author} {\bibfnamefont
  {H.}~\bibnamefont {Xu}},\ }\href {https://doi.org/10.1016/j.fuel.2020.119516}
  {\bibfield  {journal} {\bibinfo  {journal} {Fuel}\ }\textbf {\bibinfo
  {volume} {287}},\ \bibinfo {pages} {119516} (\bibinfo {year}
  {2021})}\BibitemShut {NoStop}%
\bibitem [{\citenamefont {Nasiri}\ and\ \citenamefont
  {Luo}(2017)}]{nasiri2017specificity}%
  \BibitemOpen
  \bibfield  {author} {\bibinfo {author} {\bibfnamefont {R.}~\bibnamefont
  {Nasiri}}\ and\ \bibinfo {author} {\bibfnamefont {K.~H.}\ \bibnamefont
  {Luo}},\ }\href {https://doi.org/10.1038/s41598-017-05160-z} {\bibfield
  {journal} {\bibinfo  {journal} {Sci. Rep.}\ }\textbf {\bibinfo {volume}
  {7}},\ \bibinfo {pages} {1} (\bibinfo {year} {2017})}\BibitemShut {NoStop}%
\bibitem [{\citenamefont {Weiss}\ \emph {et~al.}(2020)\citenamefont {Weiss},
  \citenamefont {Giddey}, \citenamefont {Meyer},\ and\ \citenamefont
  {Jenny}}]{weiss2020evaporating}%
  \BibitemOpen
  \bibfield  {author} {\bibinfo {author} {\bibfnamefont {P.}~\bibnamefont
  {Weiss}}, \bibinfo {author} {\bibfnamefont {V.}~\bibnamefont {Giddey}},
  \bibinfo {author} {\bibfnamefont {D.~W.}\ \bibnamefont {Meyer}},\ and\
  \bibinfo {author} {\bibfnamefont {P.}~\bibnamefont {Jenny}},\ }\href
  {https://doi.org/10.1063/5.0013326} {\bibfield  {journal} {\bibinfo
  {journal} {Phys. Fluids}\ }\textbf {\bibinfo {volume} {32}},\ \bibinfo
  {pages} {073305} (\bibinfo {year} {2020})}\BibitemShut {NoStop}%
\bibitem [{\citenamefont {Bukhvostova}\ \emph {et~al.}(2016)\citenamefont
  {Bukhvostova}, \citenamefont {Kuerten},\ and\ \citenamefont
  {Geurts}}]{bukhvostova2016heat}%
  \BibitemOpen
  \bibfield  {author} {\bibinfo {author} {\bibfnamefont {A.}~\bibnamefont
  {Bukhvostova}}, \bibinfo {author} {\bibfnamefont {J.~G.~M.}\ \bibnamefont
  {Kuerten}},\ and\ \bibinfo {author} {\bibfnamefont {B.~J.}\ \bibnamefont
  {Geurts}},\ }\href {https://doi.org/10.1016/j.ijheatfluidflow.2016.04.007}
  {\bibfield  {journal} {\bibinfo  {journal} {Int. J. Heat Fluid Flow}\
  }\textbf {\bibinfo {volume} {61}},\ \bibinfo {pages} {256} (\bibinfo {year}
  {2016})}\BibitemShut {NoStop}%
\bibitem [{\citenamefont {Abramzon}\ and\ \citenamefont
  {Sirignano}(1989)}]{abramzon1989droplet}%
  \BibitemOpen
  \bibfield  {author} {\bibinfo {author} {\bibfnamefont {B.}~\bibnamefont
  {Abramzon}}\ and\ \bibinfo {author} {\bibfnamefont {W.~A.}\ \bibnamefont
  {Sirignano}},\ }\href {https://doi.org/10.1016/0017-9310(89)90043-4}
  {\bibfield  {journal} {\bibinfo  {journal} {Int. J. Heat Mass Transfer}\
  }\textbf {\bibinfo {volume} {32}},\ \bibinfo {pages} {1605} (\bibinfo {year}
  {1989})}\BibitemShut {NoStop}%
\bibitem [{\citenamefont {Bukhvostova}\ \emph {et~al.}(2014)\citenamefont
  {Bukhvostova}, \citenamefont {Russo}, \citenamefont {Kuerten},\ and\
  \citenamefont {Geurts}}]{bukhvostova2014comparison}%
  \BibitemOpen
  \bibfield  {author} {\bibinfo {author} {\bibfnamefont {A.}~\bibnamefont
  {Bukhvostova}}, \bibinfo {author} {\bibfnamefont {E.}~\bibnamefont {Russo}},
  \bibinfo {author} {\bibfnamefont {J.~G.~M.}\ \bibnamefont {Kuerten}},\ and\
  \bibinfo {author} {\bibfnamefont {B.~J.}\ \bibnamefont {Geurts}},\ }\href
  {https://doi.org/10.1016/j.ijmultiphaseflow.2014.03.004} {\bibfield
  {journal} {\bibinfo  {journal} {Int. J. Multiph. Flow}\ }\textbf {\bibinfo
  {volume} {63}},\ \bibinfo {pages} {68} (\bibinfo {year} {2014})}\BibitemShut
  {NoStop}%
\bibitem [{\citenamefont {Wang}\ \emph {et~al.}(2021)\citenamefont {Wang},
  \citenamefont {{Dalla Barba}},\ and\ \citenamefont {Picano}}]{wang2021}%
  \BibitemOpen
  \bibfield  {author} {\bibinfo {author} {\bibfnamefont {J.}~\bibnamefont
  {Wang}}, \bibinfo {author} {\bibfnamefont {F.}~\bibnamefont {{Dalla
  Barba}}},\ and\ \bibinfo {author} {\bibfnamefont {F.}~\bibnamefont
  {Picano}},\ }\href {https://doi.org/10.1016/j.ijmultiphaseflow.2021.103567}
  {\bibfield  {journal} {\bibinfo  {journal} {Int. J. Multiph. Flow}\ }\textbf
  {\bibinfo {volume} {137}},\ \bibinfo {pages} {103567} (\bibinfo {year}
  {2021})}\BibitemShut {NoStop}%
\bibitem [{\citenamefont {{Dalla Barba}}\ and\ \citenamefont
  {Picano}(2018)}]{dalla2018clustering}%
  \BibitemOpen
  \bibfield  {author} {\bibinfo {author} {\bibfnamefont {F.}~\bibnamefont
  {{Dalla Barba}}}\ and\ \bibinfo {author} {\bibfnamefont {F.}~\bibnamefont
  {Picano}},\ }\href {https://doi.org/10.1103/PhysRevFluids.3.034304}
  {\bibfield  {journal} {\bibinfo  {journal} {Phys. Rev. Fluids}\ }\textbf
  {\bibinfo {volume} {3}},\ \bibinfo {pages} {034304} (\bibinfo {year}
  {2018})}\BibitemShut {NoStop}%
\bibitem [{\citenamefont {Ciottoli}\ \emph {et~al.}(2020)\citenamefont
  {Ciottoli}, \citenamefont {Battista}, \citenamefont {Galassi}, \citenamefont
  {{Dalla Barba}},\ and\ \citenamefont {Picano}}]{ciottoli2020direct}%
  \BibitemOpen
  \bibfield  {author} {\bibinfo {author} {\bibfnamefont {P.~P.}\ \bibnamefont
  {Ciottoli}}, \bibinfo {author} {\bibfnamefont {F.}~\bibnamefont {Battista}},
  \bibinfo {author} {\bibfnamefont {R.~M.}\ \bibnamefont {Galassi}}, \bibinfo
  {author} {\bibfnamefont {F.}~\bibnamefont {{Dalla Barba}}},\ and\ \bibinfo
  {author} {\bibfnamefont {F.}~\bibnamefont {Picano}},\ }\href
  {https://doi.org/10.1007/s10494-020-00200-7} {\bibfield  {journal} {\bibinfo
  {journal} {Flow Turbul. Combust.}\ ,\ \bibinfo {pages} {1}} (\bibinfo {year}
  {2020})}\BibitemShut {NoStop}%
\end{thebibliography}%
\end{document}